\pdfoutput=1
\documentclass{article}

\usepackage[final]{neurips_2025}

\usepackage{color}
\usepackage[dvipsnames]{xcolor}
\usepackage{colortbl}
\usepackage[utf8]{inputenc} 
\usepackage[T1]{fontenc}    

\usepackage[colorlinks,
            linkcolor=VioletRed,      
            anchorcolor=VioletRed, 
            citecolor=VioletRed,       
            ]{hyperref}
\usepackage{url}            
\usepackage{booktabs}       
\usepackage{amsfonts}       
\usepackage{nicefrac}       
\usepackage{microtype}      
\usepackage{xcolor}         
\usepackage{xfrac}
\usepackage{authblk}
\usepackage{amsmath}
\usepackage{cleveref}
\usepackage{multirow}
\usepackage{multicol}
\usepackage{colortbl}
\usepackage{bm}
\usepackage{float}
\usepackage{graphicx}
\usepackage{caption}
\usepackage{subcaption}
\usepackage{tabularx}

\newcommand{\M}{{KuaiMod}}
\newcommand{\C}{{Kuaishou}}
\newcommand{\Y}{{YuanQi}}
\title{VLM as Policy: Common-Law Content Moderation Framework for Short Video Platform}
\author[1,2${\dagger}$]{Xingyu Lu}
\author[1${\dagger}$]{Tianke Zhang}
\author[1]{Chang Meng}
\author[1]{Xiaobei Wang}
\author[2]{Jinpeng Wang}
\author[3]{Yi-Fan Zhang}
\author[1]{Shisong Tang}
\author[1]{Changyi Liu}
\author[1]{Haojie Ding}
\author[1]{Kaiyu Jiang}
\author[1]{Kaiyu Tang}
\author[1]{Bin Wen}
\author[2]{Hai-Tao Zheng}
\author[1${\ddagger}$${\spadesuit}$]{Fan Yang}
\author[1]{Tingting Gao}
\author[1]{Di Zhang}
\author[4]{, Kun Gai}
\affil[ ]{{$^1$Kuaishou, $^2$Tsinghua, $^3$CASIA, $^4$Unaffiliated}}

\renewcommand{\paragraph}[1]{\medskip\noindent\textbf{#1.~}}

\newcommand{\bma}{\bm{a}}

\newcommand{\bmt}{\bm{t}}
\newcommand{\bmu}{\bm{u}}
\newcommand{\bmv}{\bm{v}}

\newcommand{\bmL}{\bm{L}}

\begin{document}
\maketitle
\vspace{-1.2cm}
\begin{center}
$^{\dagger}$ Equal Contribution
$^{\ddagger}$ Corresponding Author 
$^{\spadesuit}$ Project Leader \\
\url{https://kuaimod.github.io}
\end{center}

\begin{abstract}
Exponentially growing short video platforms (SVPs) face significant challenges in moderating content detrimental to users' mental health, particularly for minors. The dissemination of such content on SVPs can lead to catastrophic 
societal consequences. Although substantial efforts have been dedicated to moderating such content, existing methods suffer from critical limitations: (1) \textbf{Manual review} is prone to human bias and incurs high operational costs.  
(2) \textbf{Automated methods}, though efficient, lack nuanced content understanding, resulting in lower accuracy.  
(3) \textbf{Industrial moderation regulations} struggle to adapt to rapidly evolving trends due to long update cycles. In this paper, we annotate the first SVP content moderation benchmark with authentic user/reviewer feedback to fill the absence of benchmark in this field. Then we evaluate various methods on the benchmark to verify the existence of the aforementioned limitations.
We further propose our \textbf{common-law} content moderation framework named \M{} to address these challenges. \M{} consists of three components: training data construction, offline adaptation, and online deployment \& refinement.
Leveraging large vision language model (VLM) and Chain-of-Thought (CoT) reasoning, \M{} adequately models video toxicity based on sparse user feedback and fosters dynamic moderation policy with rapid update speed and high accuracy.
Offline experiments and large-scale online A/B test demonstrates the superiority of \M{}: \M{} achieves the best moderation performance on our benchmark. The deployment of \M{} reduces the  user reporting rate by 20\% and its application in video recommendation increases both Daily Active User (DAU) and APP Usage Time (AUT) on several \C{} scenarios. 
We have open-sourced our benchmark at \href{https://github.com/KuaiMod/KuaiMod.github.io}{https://github.com/KuaiMod/KuaiMod.github.io}.

\end{abstract}


\section{INTRODUCTION}
\label{sec: introduction}
Short video platforms (SVPs) such as \C{} and TikTok  have experienced exponential growth in the Internet era, delivering millions of new videos daily. 
Videos of high quality delivery knowledge and joy to users and ensure the social value of SVPs. Meanwhile, widespread harmful videos pose significant threats to the ecosystem of the platform \cite{spread_1, minors}, highlighting the need to develop a robust and effective ecosystem governance system \cite{jigsaw}. 

To prevent inappropriate content from harming the platform ecosystem, the critical content moderation stage \cite{def1} has emerged to preemptively filter such content. 
The content moderation workflow on SVPs involves the systematic evaluation of videos to ensure their compliance with legal regulations, platform policies, and social ethics \cite{def2,def3}. 
The traditional SVP content moderation paradigm follows the \textbf{civil law system} \cite{civil}: Platforms establish a rule-based system grounded in laws and social consensus, while annotators act as judges to first understand the rule system then traverse to determine videos' harmfulness. This paradigm suffers from mental toll to annotators, labor costs and susceptibility to human bias \cite{youtube}.

\begin{figure*}
    \centering
    \begin{subfigure}[b]{0.57\linewidth}
        \centering
        \includegraphics[width=\linewidth, height=5.5cm]{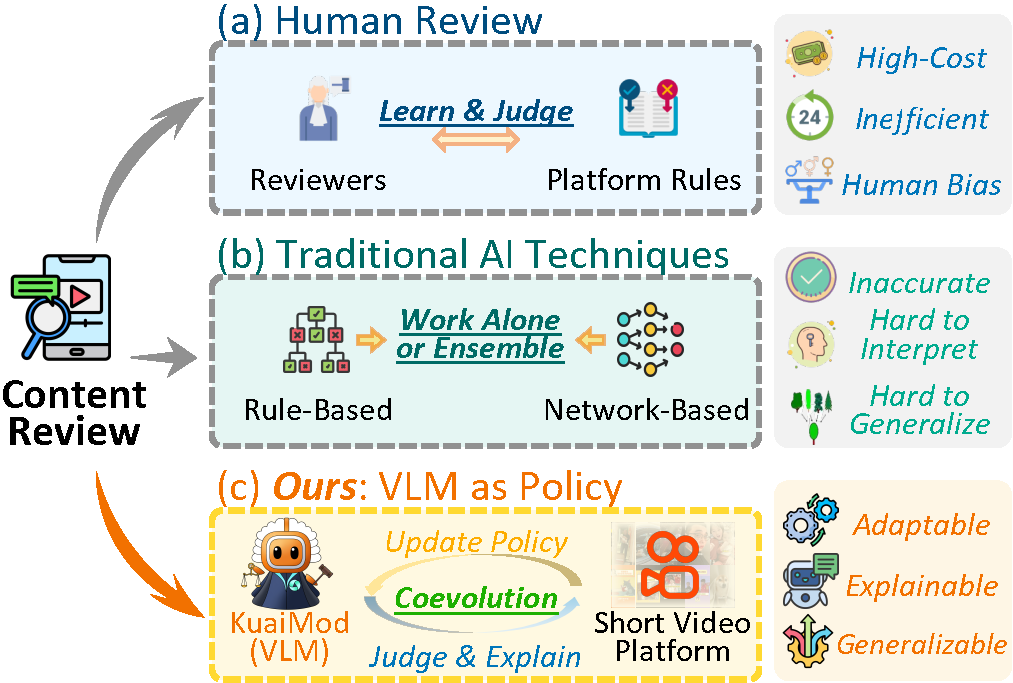}
        \caption{A visualization about three moderation paradigms.}
        \label{fig:intro}
    \end{subfigure}
    \hfill
    \begin{subfigure}[b]{0.39\linewidth}
        \centering
        \includegraphics[width=\linewidth, height=5.5cm]{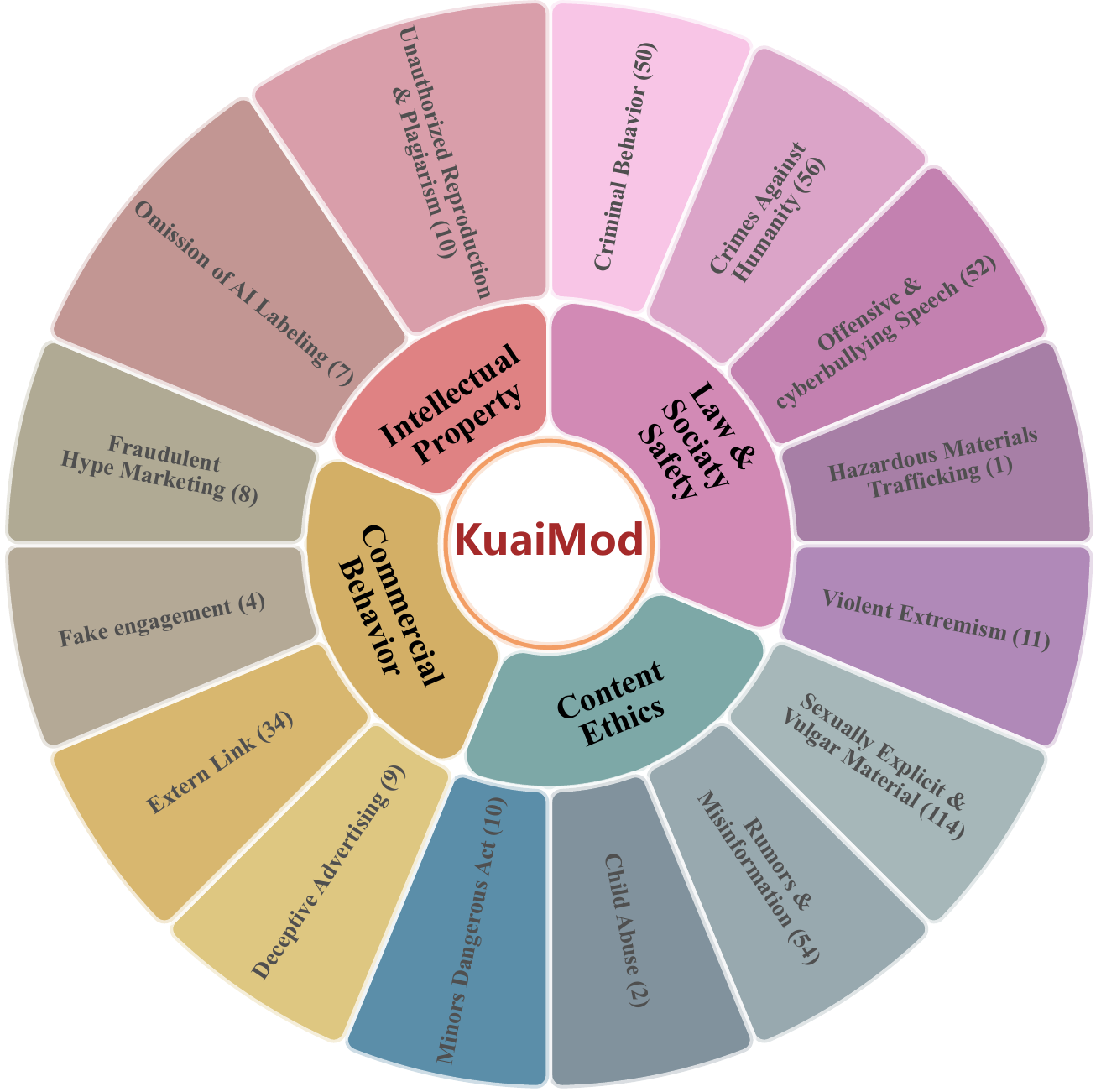}
        \caption{An overview of \M{} benchmark.}
        \label{fig:bench}
    \end{subfigure}
    \label{fig:combined}
    \vspace{-0.5cm}
\end{figure*}

Recently, artificial intelligence technologies are employed to reduce manual workload in content moderation with automated workflows.
Early methods either encode rules, such as keyword matching \cite{keyword}, into programs or train classifiers, including decision tree \cite{decision_tree}, Naive Bayesian \cite{bys}, and random forest \cite{automated3}, on labeled data to filter harmful content. These methods suffer from unsatisfactory accuracy \cite{awful2} and are only effective in specific scenarios.
More recent advances leverage large language models (LLMs) with exceptional content understanding capabilities to assist \cite{adapt_LLM} or replace human reviewers \cite{adapt_LLM}. For instance, Google's Jigsaw API \cite{jigsaw} applies a pre-trained LLM to assess the harmfulness of content. Other approaches prompt LLMs to moderate and neutralize harmful content \cite{prompt_once}. Additionally, some studies apply the divergences between LLM outputs and human review to refine moderation policy \cite{update_policy}. 

To validate the performance of these moderation methods under the SVP scenario, we construct a benchmark based on user feedback and manual review from the \C{} platform, which contains 422 videos classified as harmful.
Our evaluation result reveals that:  
Despite the advancements of existing methods, several unique challenges for SVP content moderation are remain unresolved:

\begin{itemize}
    \item \textbf{Complex Content Understanding:} For effective moderation on SVPs, it is essential to understand video content and user demands. However, early solutions and small-scale classifiers are unable to grasp the semantic of videos, while LLM-based methods struggle to process complex multi-modal content, let alone evaluating a video from user perspective.
    \item \textbf{Rapidly Evolving Trends:} Driven by unpredictable societal events, trends on SVPs evolve so rapidly that static rule-based and model-based systems are prone to be incomplete. Unfortunately, prolonged update cycles further hinder these systems in detecting and preventing the dissemination of newly emerging harmful content.
    \item \textbf{Deployment Requirements:} Industrial SVPs typically involve millions of active daily users and newly uploaded videos, placing stringent demands on the moderation accuracy and deployment costs. Previous research lacks industrial deployment experience and has not demonstrated moderation effectiveness on real-world SVPs.
\end{itemize}

This study draws inspiration from the \textbf{common law system} \cite{common} to address these challenges. Common law system relies on existing cases rather than static legal statutes to judge whether the defendant is violative. Compared to civil law, the common law system offers greater flexibility: Case-based precedents represent a consensus and continuously evolve with social progress. Emulating the common law system, we develop a case-based framework named \M{} for SVP content moderation. 

\M{} consists of three components: 
(1) \textbf{Data Construction}: We extract positive and violative examples from \C{} platform to construct chain-of-thought reasoning data. All instances contain detailed rationales and a verdict as reference for video moderation.
(2) \textbf{Offline Adaptation}: Instead of static regulations, we transform \C{} self-developed \Y{} VLM \cite{yuanqi} into dynamic case-based moderator through curriculum training. 
(3) \textbf{Online Deployment \& Refinement}: Our deployed VLM judge provides fine-grained judgment for each video and refines its policy according to online feedback. These judgments are further used to enhance online personalized video recommendation.

As visualized in \Cref{fig:intro}, \M{} specifically overcomes the aforementioned challenges with targeted strengths: 
 (1) \textbf{Comprehensive understanding}: Training paradigm with video CoT data reinforces the powerful VLM to better interpret heterogeneous video metadata including titles, frame images and user comments.
 (2) \textbf{Rapid updatability}: Our case-based moderation policy effortlessly achieves a daily update frequency: Since all processes are reusable, policy updates require only the addition of new violations and model training.
 (3) \textbf{High deployability}: The VLM judge, with an appropriate parameter scale, balances high accuracy and low inference cost, enabling moderation over millions of videos per day. 
We summarize our contributions as follows:

\begin{itemize}
    
    \item \textbf{First SVP Moderation Benchmark}: Recognizing the importance of harmful content moderation for SVPs, we constructed a content moderation benchmark with authentic user feedback from \C{} platform and strict data annotation. After removing privacy and illegal content, we open-source it to fill a critical gap in this field. To our knowledge, \M{} is the first SVP content moderation benchmark.
    
    \item \textbf{Common-Law Moderation Framework}: An evaluation on our SVP governance benchmark reveals several limitations of existing moderation methods. To address these drawbacks, we propose our video content moderation framework named \M{}. \M{} can dynamically update governance policy by tracking immediate user feedback, and achieves accuracy comparable to human review, outperforming all baselines.
    
    \item \textbf{Industrial Deployment \& Validation}: We demonstrate the significant industrial value of \M{}: In several \C{} scenarios with tens of millions of users, A/B tests show that \M{} achieves a 20\% reduction in video reporting rates. When applied for personalized recommendation, \M{} achieves positive gains in DAU (Daily Active Users). \M{} has been fully deployed, serving hundreds of millions of users on \C{}.
\end{itemize}

\section{RELATED WORKS}
\label{sec: related_works}

\subsection{Content Moderation for Online Platform}
Toxic content \cite{first_hate, second_hate, comment_hate} like hate speech has been a long-standing problem for the safety of online platforms \cite{web_hate,long_exist_1, long_exist_2} with large user scale. To prevent the spread of these content \cite{spread_1}, both manual \cite{crowdsource1, crowdsource2} and automated approaches \cite{automated1, automated2} have been developed.

Traditional methods rely on crowd-sourcing \cite{crowdsource1, crowdsource3} to assess the harmfulness of media content. Manual methods typically involve creating a taxonomy and rule-based policy \cite{crowdsource1} for violations, with reviewers providing decisions accordingly. Although effective, these approaches reveals issues including high emotional and economic burdens \cite{crowdsource2}, as well as reviewer bias \cite{bias}. Therefore, automated methods are sought to address these shortcomings \cite{automated3}.

To compensate for the limitations of manual moderation, automated content moderation has become a research focus in both academia and industry \cite{keyword} in recent years, aiming to address the growing issue of harmful content on online platforms.  
Major companies have made significant progress in this field. For example, Jigsaw and Google \cite{jigsaw} developed Perspective API, a machine learning-based tool designed to detect and assess the toxicity of online comments. 
Twitter employs machine learning models to identify and remove hate speech and other harmful content \cite{twitter1, twitter2}. These models analyze tweet text, user behavior, and contextual information, providing an automated approach to content moderation. 
Alibaba \cite{alibaba,alibaba2} has developed an intelligent content moderation system that leverages machine learning and deep learning techniques to automatically detect and handle inappropriate content on its e-commerce platforms. However, most of these methods rely on rule-based approaches and small-scale models, requiring coexistence with manual moderation. They suffer from significant latency and weak multi-modal understanding \cite{awful}, making them less adaptable to the fast-evolving, information-rich landscape of the modern internet environment \cite{moderation_survey1,moderation_survey2}.

\subsection{Language Models for Content Moderation}
With advancements in natural language processing (NLP) technology \cite{gpt4,deepseek}, large language models (LLMs) have become increasingly prevalent in the automated content moderation of social media platforms.
In recent years, Transformer-based LLMs, such as BERT \cite{bert} and GPT \cite{gpt}, have achieved remarkable success in natural language processing (NLP) tasks through self-attention mechanisms and large-scale pre-training. These models can effectively capture complex semantic relationships in text \cite{transformer}. In the context of content moderation, LLMs have been utilized to detect inappropriate speech, hate speech, and misinformation \cite{misinformation1,hatespeech1}.

LLMs can be adapted to specific content moderation tasks through fine-tuning. For example, by fine-tuning on datasets labeled with harmful or inappropriate content, these models can enhance their ability to detect specific types of undesirable content \cite{ULMFiT,xlnet}. Moreover, some LLMs like GPT-4o \cite{4o}, possess zero-shot and few-shot learning capabilities, enabling them to perform content moderation tasks without dedicated training data. This ability allows models to rapidly adapt to emerging content moderation needs without requiring large-scale annotated datasets \cite{fewshotcls,fewshotlearner}.

Beyond text-based models, recent research has explored the use of vision-language models (VLMs) with multi-modal understanding capabilities for moderating multi-modal content \cite{multimodel1,multimodel2}. However, these approaches have limitations in terms of generalization, particularly when dealing with diverse and complex malicious content, where models may struggle to accurately identify emerging misinformation \cite{multimodel_misinform}. Additionally, although existing LM-based moderation models possess strong analytical capabilities, they suffer from poor real-time performance, making it challenging to meet the demands of high-frequency content moderation, resulting in delays and processing bottlenecks \cite{moderation_detect_1,moderation_detect_2}.


\section{BENCHMARK}
\label{sec: benchmark}

In this section, we first formally define the video moderation task, then introduce our taxonomy for violative content on \C{} platform and \M{} benchmark based on the taxonomy.

\subsection{Task Definition for Video Moderation}
\label{subsec: task_def}
Given a video's metadata, including video frames, description text, audio, and user comments, the moderation task can be formulated as a mapping process from these inputs to a specific video label ${\bm{l}}$. We formalize the video moderation method as a mapping function:
\begin{gather}
\mathbf{f_m}(\mathbf{v}, \mathbf{t}, \mathbf{a}, \mathbf{u}) = \mathbf{l},\quad\mathbf{l} \in \mathbf{L},
\end{gather}
where $\bmv$, $\bmt$ and $\bma$ represent visual, textual and audio information respectively, $\bmu$ represents {user comments}, and $\bmL$ is the set of all labels. The output of different methods can be text or probabilities.

The {simplified version} of the video moderation task only requires {binary classification}, where $ \bmL $ consists of two labels: {positive (non-violative)} and {violative}.
However, {\M{} moderation paradigm} aims for {fine-grained classification}, categorizing videos into 1 positive label and 15  violative labels. This {granular classification} improves the {interpretability} of the moderation process and provides a structured reference for {downstream applications}.


\subsection{\M{} Taxonomy} 
\label{sec:taxonomy}
\M{} taxonomy categorizes violative content into 4 general categories: \textit{Social Security \& Law}, \textit{Content Ethics}, \textit{Prohibited Commercial Behavior} and \textit{Intellectual Property}. Each primary category is further divided into 2-5 subcategories, resulting in a total of 15 subcategories, as shown in \Cref{fig:bench}. 

The taxonomy originates from \C{}’s manual moderation guidelines, which are structured similarly to the civil-law system, defining all types of violative content with exhaustive enumeration and detailed descriptions. 
In contrast, our taxonomy is more concise and generalized to enhance the generalizability of \M{}, making it more flexible and responsive for SVP content moderation. 

Moreover, the taxonomy itself is dynamic and is progressively refined in following data construction and training process (\Cref{subsec:stage1}), making \M{} significantly easier to update compared to cumbersome regulations. This adaptability ensures the robustness of \M{} in handling evolving violative content. Note that the online version of \M{} taxonomy has been continuously expanded, now it includes over 100 labels and serves as the reference for \C{} video moderation. For simplicity in evaluation, the taxonomy presented here corresponds to its initial version.  
\subsection{\M{} Benchmark} 
\label{sec:benchmark_construction}
Based on this taxonomy, we construct our \M{} content moderating benchmark. The benchmark consists of 1,000 short videos from the \C{} platform, encompassing both healthy positive content and all 15 types of violative content.

Adhering to \C{}'s \textbf{User Obsession} principle, we collect both positive and violative content based on real user feedback. The positive instances come from the \textbf{High Quality Videos}, which are videos from the manual reviewed daily high-quality video queue that have not been reported by users and have a dislike ratio of less than 0.1\%. For violative content, there are three main sources:

\begin{itemize}
    \item \textbf{High Report Videos}: The top-$n$ videos among the daily uploaded videos, sorted by the number of reports, with each video receiving more than 30 reports.
    \item \textbf{High Dislike Videos}: The top-$n$ videos among the daily uploaded videos, sorted by the percentage of users clicking dislike, with each video's dislike rate above 50\%.
    \item \textbf{Negative Comment Videos}: Videos with more than 2,000 plays that negative comments accounting for more than 80\% of its total comments among the daily uploaded videos.
\end{itemize}

With various samples collected from the above sources, we conduct manual annotation with crowd-sourcing. 30 experienced annotators first determine whether each video is positive or violative. For violative videos, annotators further assign fine-grained labels. Finally, we select 1,000 samples with consistent annotations from all labeled data to construct the test set that covers all 15 violative categories. Each sample in \M{} benchmark contains multiple frames extracted from the video, along with textual content, including title, text recognized via OCR and ASR, and user comments.  

To the best of our knowledge, \M{} is the first open-source benchmark for industrial SVP content moderation. We believe it holds significant value in advancing SVP ecosystem governance.

\section{METHODOLOGY}
\label{sec:method}

In this section, we present how to develop the \M{} moderation model based on our self-developed general VLM, \Y{}. We first introduce \Y{} in \Cref{sec: yuanqi}, then we explain the data construction, training, deployment and updating pipeline of \M{}.  
\Cref{subsec:train_data_construction} presents the data construction process: We utilize largest \Y{}-20B model to construct moderation data in a state-transition format with CoT technology.  
\Cref{subsec:stage1} details the post-training process of \M{}, which adapts the general \Y{}-7B model into a video moderation model with Supervised Fine-Tuning (SFT) and Direct Preference Optimization (DPO).  
\Cref{subsec:stage2} describes the online deployment of \M{} and its RL-based updating mechanism to iteratively refine moderation policy.

\subsection{Overview of \Y{} Model}
\label{sec: yuanqi}
\subsubsection{Architecture \& Modules}

The \Y{} model is an efficient VLM following the Flamingo-style architecture, consisting of:
    (1) \textbf{Visual Encoder:} Built upon the EVA2-CLIP-E-Plus model, it extracts hierarchical vision features at multiple levels, enabling fine-grained perception of visual elements.
    (2) \textbf{LLM Foundation:} Based on Qwen-14B-Chat 1.0, providing strong content understanding and logical reasoning capabilities.
    (3) \textbf{Gated Cross-Attention Layers:} Inserted before each transformer layer in the LLM, allowing effective image-text fusion while maintaining high computational efficiency.
    (4) \textbf{Mixture of Experts (MoE):} Optimizes model effectiveness by dynamically activating specialized subnetworks.

\subsubsection{Training}
The \Y{} model undergoes three-stage training:
    (1) \textbf{Multi-Modal Pre-Training} aligns visual and textual modalities with large-scale image-text pairs, ensuring high relevance through extensive data cleaning.
    (2) \textbf{Multi-Task Continual Pre-Training} incorporates tasks including visual question answering (VQA), OCR, and object detection, enhancing high-level reasoning capabilities.
    (3) \textbf{Supervised Fine-Tuning} activates instruction-following abilities with high-quality instruction data.
\subsubsection{Performance \& Application}

The hierarchical feature extraction mechanism, cross-attention architecture, and multi-stage training pipeline enable \Y{} to achieve state-of-the-art performance on diverse benchmarks including general  VQA, fine-grained visual perception, bilingual capabilities and hallucination detection.
Overall,  \Y{} is highly suitable for tasks requiring complex visual and textual integration, particularly in short video scenarios.
Additionally, the \Y{} model scales from 3B to 20B. In our approach, we use the largest 20B version for data generation due to its superior inference and instruction-following capabilities, while the smaller 7B version serves as the foundation for \M{} moderation model, ensuring efficacy-efficiency balance for deployment.  

\subsection{Construction of Training Data}
\label{subsec:train_data_construction}

\subsubsection{{Data Collection and Annotation}} 
The data source and labeling process for training data is the same as the benchmark construction in \Cref{sec:benchmark_construction}, but at a larger scale. We sampled 50,000 videos from four video queues on the Kuaishou platform between May 1st and November 1st—a six-month period preceding the project launch—for training data construction, ensuring data freshness.  

Regarding the annotation process, unlike the static taxonomy shown in \Cref{fig:bench}, the taxonomy for training data is continuously expanded based on annotation results, as previously mentioned: Experienced moderators are first required to determine whether a video violates platform policies. For violating videos, they assign them to a primary violation category and then assess whether the violation type matches an existing secondary label. For cases that cannot be classified into the current taxonomy by human annotators, we instruct the \Y{}-20B model to generate supplementary secondary labels. We then collect all newly generated secondary labels under each primary category, aggregate them, and manually review before incorporating them into the original taxonomy.  

In this way, we have expanded the taxonomy to include over 100 violation categories and continue to extend it after online deployment. 
At the end of the annotation process, each video is assigned a violation type tag, which serves as the basis for the subsequent data construction process. The initial taxonomy covers 24,562 cases, we take this split as training data for offline evaluation.
\subsubsection{Generate State-Transition Data with Tag2CoT and CoT2Tag}
\label{subsec: workflow}
Instead of directly using the tag as the training target, we aim to model the causality between video content and the manual moderation result. Specifically, given a video's meta data and its violation type tag as input context, we prompt \Y{}-20B model with instructions to analyze the video's content then generate Chain-of-Thought rationales to explain why the video is classified as its violation tag. This improves the interpretability of the moderation result. We refer to this process as \textbf{Tag2CoT}.

\textbf{Tag2CoT} process produces causal analysis between metadata and violation tags.  
However, as described in \Cref{subsec: task_def}, The video moderation is a process of mapping video metadata to violation labels, the rationales generated in \textbf{Tag2CoT} process are insufficient for training a VLM to complete the moderation task since its analysis proceeds from the outcome to the cause. Therefore, we introduce a second process, \textbf{CoT2Tag}, to organize these elements into a video moderation workflow. 

We first explain our workflow for video moderation: We introduce the concept of state transition from reinforcement learning (RL) to analyze videos in a structured, step-by-step manner. The initial state is the input context, which contains the task definition and video metadata. Next, the VLM sequentially transits to the following states and ultimately makes its moderation decision.  

(1) \textbf{Content Extraction}: In this state, VLM extracts content from video frames and generates an overall description of the video, converting all metadata into textual format.  

(2) \textbf{Video Content Analysis}: In this state, VLM analyzes all the meta data including the video description, title, OCR, and ASR transcripts sequentially to identify any explicit violations.  

(3) \textbf{Interim Check}: In this state, VLM determines whether a violation category can be assigned based on the previous analysis. If the decision remains uncertain, proceed with further examination.  

(4) \textbf{User Feedback Analysis}: In this state, for cases that are difficult to determine based on metadata alone, VLM incorporates user comments to assess the video's impact on viewers and finally decides the video's violation category.  

(5) \textbf{Final Summary}: In this state, VLM summarizes the findings from all previous states, assigns the final violation tag, and provides several key reasons as support for its judgment.  

In the \textbf{CoT2Tag} process, our \Y{}-20B model takes the video metadata, the moderation tag and rationales as input and convert them into the state-transition format data. Each state performs causal analysis about a specific aspect of the video, and the last state provides the moderation result.
A more intuitive representation of the workflow can be found in the \Cref{fig:case1}. 

For the state transition data generated by \Y{}-20B, we first prompt \Y{}-20B to self-verify the correctness of its output then conduct manual review to identify logically inconsistent or misattributed samples then correct them. These bad cases serve as negative samples for model training.  
It is important to note that while human annotation plays a crucial role, the majority work in our data construction process is handled by the \Y{} model. Additionally, the scale of human annotation is negligible compared to the platform's daily moderation demands. Besides, these annotated cases can be reused, further reducing costs.

\subsection{Stage I: Offline Adaptation Training}
\begin{figure*}[h]
    \centering
    \includegraphics[width=\textwidth]{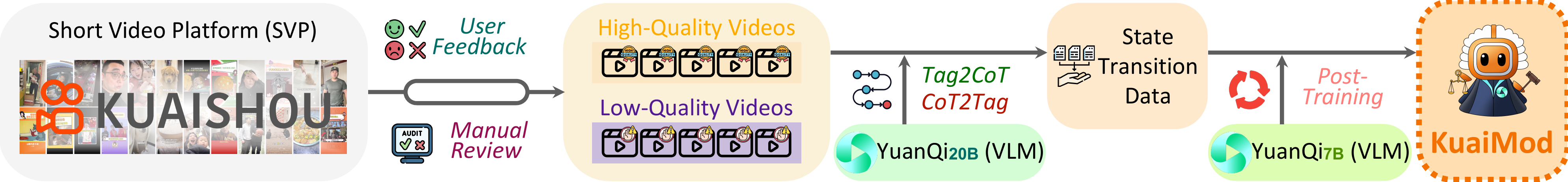}
    \caption{Offline adaptation stage of \M{}: We post-train the \Y{}-7B with state-transition format data. After SFT and DPO, \M{} is transformed into a video moderator for online services. 
}
    \label{fig:stag1}
\end{figure*}

As shown in \Cref{fig:stag1}, in offline adaptation stage, we adapt the general \Y{}-7B model to \M{} video moderator via two training steps: Large-scale SFT and mistake-oriented DPO.
\label{subsec:stage1}
\subsubsection{Large-scale SFT} We take all the right cases as corpus to SFT the \Y{}-7B model. The SFT loss is formulated as:
\begin{equation}
\mathcal{L}_{\text{SFT}} = - \frac{1}{N} \sum_{t=1}^{N} \log P_{\theta}(y_t |x,y_{<t}),
\end{equation}
$N$ denotes the number of output tokens, $y_t$ represents the $t$-th token in the target response, 
$y_{<t} = (y_1, y_2, \dots, y_{t-1})$ denotes the first $t-1$ output tokens, $x$ denotes the input tokens, including video frames, title, OCR/ASR text and user comments as mentioned in \Cref{subsec: task_def}.
$P_{\theta} (y_t{\mid}y_{<t})$ represents the conditional probability distribution over $y_t$ given $y_{<t}$, where $\theta$ denotes the model parameters.  
$\log P_{\theta} (y_t{\mid}y_{<t})$ is the logarithm of the probability from the \textit{softmax} function.  

At this stage, our VLM learns from a large-scale cases about how to analyze video content through the state transition process and complete the video moderation task.  
Adopting a divide-and-conquer strategy, decoupling the analysis of different aspects of the video into independent states before aggregating the results, not only prevents overlook of important details but also enables a more comprehensive assessment of the video's harmfulness.  

\subsubsection{Mistake-oriented DPO}
In this step, we leverage DPO to correct the model's mistakes and improve moderation accuracy.

After several epochs of SFT, we take the trained model to inference on the SFT dataset. On each data instance, we sample multiple times to collect correct/incorrect responses that align/misalign with the reference answer, then use them to construct paired samples for DPO training, where each pair consists of one correct output and one incorrect output. 
Additionally, during the data generation process, we corrected some erroneous instances. These corrected and incorrect samples are also paired for DPO training.  
The DPO loss is formulated as:
\begin{equation}
\begin{aligned}
\label{DPO}
\mathcal{L}_{\mathrm{DPO}}(\pi_{\theta};\pi_{\mathrm{ref}}) 
= -\mathbb{E}_{(x, y_r, y_w) \sim \mathcal{D}} \Bigg[\log \sigma \Bigg( \beta \log \frac{\pi_{\theta}(y_r{\mid}x)}{\pi_{\mathrm{ref}}(y_r{\mid}x)} 
- \beta \log \frac{\pi_{\theta}(y_w{\mid} x)}{\pi_{\mathrm{ref}}(y_w{\mid}x)}
\Bigg) \Bigg],
\end{aligned}
\end{equation}
in which $y_r$ and $y_w$  represents the correct and incorrect response respectively,
$\pi_{\theta}(y{\mid}x)$ denotes the probability assigned by the current model with parameters $\theta$ to response $y$ given input $x$, whereas $ref$ represents the reference model, which is the SFT model.  
The scaling factor $\beta$ controls the strength of preference optimization. The term  
$\log \frac{\pi_{\theta}(y_r \mid x)}{\pi_{\mathrm{ref}}(y_r \mid x)}$ and $\log \frac{\pi_{\theta}(y_w \mid x)}{\pi_{\mathrm{ref}}(y_w \mid x)}$
capture the log-probability ratio of the correct and incorrect response between the current and reference models.
Finally, the sigmoid function $\sigma(\cdot)$ ensures that the loss guides the model to assign higher probabilities to correct responses while decreasing the probabilities of incorrect ones, effectively refining its moderation decisions.  

\subsection{Stage II: Online Deployment \& Refinement}
\label{subsec:stage2}
\begin{figure*}[h]
    \centering
    \includegraphics[width=\textwidth]{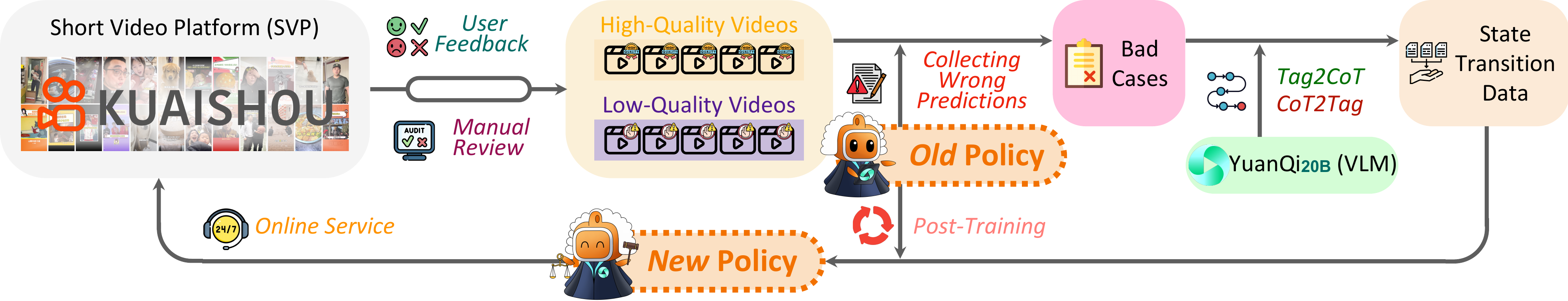}
    \caption{Online deployment stage of \M{}: The initially trained \M{} model is deployed into \C{} as a moderation agent. \M{} interacts with the online environment and iteratively refines its policy with user feedback in the RL manner.}
    \label{fig:stage2}
\end{figure*}

Although \M{} model acquires preliminary video moderation capabilities after offline adaptation, as previously analyzed, the scope of violation contents in SVPs can evolve with users and social trends. A static moderation model cannot effectively adapt to the dynamic nature of SVPs.  
Therefore, in Stage II, we design an update mechanism as depicted in \Cref{fig:stage2}, which follows Reinforcement Learning from User Feedback (RLUF) paradigm to continuously refines \M{}'s moderation policy based on SVP feedback.  


\subsubsection{RLUF Definition for \M{} Online Refinement}
We introduce key elements of the RLUF paradigm and explain their corresponding roles in \M{}'s online refinement process:

- Environment (\( \mathcal{E} \)): The environment consists of videos and users on the SVP. Videos vary in content and potential violations, while users  consume videos and provide feedback including reporting.  

- Agent (\( \mathcal{A} \)): Our \M{} model serves as an agent interacting with the environment, its policy  $\pi_{\theta}$ is parameterized by \( \theta \), which determines moderation judgments based on videos' content.  

- Reward and Objective: The reward signal is directly derived from user feedback, where a lower online user report rate indicates the better moderation policy. The objective of \M{} is to update its moderation policy through interactions with the environment, ensuring alignment with users' understanding of violation content, to improve user satisfaction and platform ecosystem.



\subsubsection{RLUF Data construction}
The key to aligning user preferences with the model's moderation policy lies in eliminating discrepancies between their understandings of violation content. Therefore, we select video cases where the model's moderation decisions deviate from user perception as training data for the refinement phase.


\textbf{Real-time Hard Case Collection.} 
We still select the four data queues in \Cref{sec:benchmark_construction} as data sources. In addition, we introduce a new queue: \textbf{High-Engagement Videos}, which refers to the most popular daily videos. These videos have a large audience, meaning that any violations within them pose a greater risk to the SVP ecosystem. Additionally, the rapidly changing nature of popular content directly reflects platform trends and user preferences, making it a critical focus for moderation.  


\textbf{Two-stage Data Curation.}
(1) \textit{Coarse filtering}. We first assume that videos from the high-quality and high-engagement queues are positive, while those sampled from the high-report, high-dislike, and negative-comment queues are violative. Then \M{} model with the old policy generates moderation decisions for all data so that we can identify cases where the model's judgment contradicts human feedback as candidates for refinement. (2) \textit{Fine-grained annotation}. Given the controversial candidates, we assign the final violation labels through human review and then generate data using the previously established Tag2CoT and CoT2Tag processes. 
For cases where \M{} makes incorrect judgments, the original response and the newly generated response can be directly used as the incorrect answer and correct answer, respectively.  
For cases where \M{}'s decision is correct but some users still perceive the content as inappropriate, we instruct the \Y{}-20B model to enumerate, within the CoT reasoning process, the aspects of the video that might cause user discomfort and explain why they do not in fact violate platform regulations. This serves as positive examples to help the model distinguish between individual user sensitivities and societal consensus.



\subsubsection{Online RLUF training}


Similarly to the mistake-oriented DPO in offline training, we still choose DPO algorithm in \cref{DPO} for online refinement training. Compared to offline training, in addition to differences in data sources, the negative samples in the online refinement phase are entirely sampled by the deployed \M{} model. And the online refinement process undergoes multiple rounds of continuous optimization to keep \M{} up with the evolving trends of \C{}. With a data scale in the hundreds, \M{} achieves low-cost daily updates.

\section{OFFLINE EVALUATION}
\label{sec: experiments}
\subsection{Experimental Setup}
\subsubsection{Dataset} Following our construction workflow in \Cref{subsec:train_data_construction}, we build a larger data set for training. This dataset contains 24,562 samples, covering all 15 types of violative videos. In \Cref{tab:data_scale}, we provide an overview of the distribution of the training and test sets. 
For all methods that requires training, including our \M{}, we use the 24,562 samples as training data to ensure fair comparison.

\begin{table}[h]
    \centering
    \vspace{-0.2cm}
    \caption{Data Distribution in Train and Test split.}
\vspace{0.1cm}
    \label{tab:data_scale}
    \begin{tabular}{lcc}
        \toprule
        \textbf{Category} & \textbf{Train} & \textbf{Test} \\
        \midrule
        Positive       & 19,343  & 578  \\
        Law \& Safety    & 1,862  & 170  \\
        Content Ethics & 2,659  & 180  \\
        Commercial     & 511   & 55   \\
        Compliance     & 187   & 17   \\
        \bottomrule
    \end{tabular}
\end{table}

\subsubsection{Metrics}

Considering that certain baselines do not support multi-class classification, and the primary target for content moderation is to distinguish positive and negative content, we focus on comparing binary classification performance. 
The utilization of multi-class classification is discussed in \Cref{recommend}

For binary classification, we employ metrics including \textit{Recall}, \textit{Precision}, and \textit{Accuracy} to evaluate the performance of various methods. We report the overall accuracy of various methods on the benchmark, along with their precision and recall for both positive and negative content, as well as the recall for four major violation categories.  We unify the judgment results across all methods, including both binary and multi-class models, by treating any prediction that is not positive as negative.  

\subsubsection{Baselines} For a comprehensive comparison, we have selected the following three groups of baselines:

\noindent{(1) Methods based on API.}
\begin{itemize}
    \item \textbf{Perspective API} \cite{jigsaw}: An industrial-grade text moderation API developed by Google. Given a text input, we utilize the returned {TOXICITY} and {SEVERE\_TOXICITY} scores to obtain harmfulness classification results  with predefined thresholds.  

    \item \textbf{GPT-4o mini \& GPT-4o API} \cite{4o}: GPT-4o mini and GPT-4o are general models of OpenAI, capable of processing multi-modal inputs. We prompt them to judge each video's category with different instructions for binary and multi-class classification.
\end{itemize}

\noindent{(2) Methods based on tiny models.}

\begin{itemize}
    \item \textbf{RoBERTa} \cite{roberta}: One of the most effective text encoders, widely used for various natural language processing tasks including Question Answering and Sentiment Analysis.  

    \item \textbf{CN-CLIP} \cite{cn_clip}: A CLIP-style image-text encoder for Chinese scenarios. We concatenate its text and image representations as input to the MLP classifier.  

\end{itemize}

\noindent{(3) Methods based on large models.}
\begin{itemize}
    \item \textbf{YOLO} \cite{prompt_once}: The first method that leverages soft prompt tuning to train LLMs for toxic content classification and text detoxification.  
    \item \textbf{Intern-VL} \cite{internvl}: One of the most advanced open-source VLMs, we directly fine-tune its 7B model for binary/multi-class classification.  

\end{itemize}

\subsubsection{Implementation Details}
For the Perspective API, we select the threshold that maximizes overall accuracy by searching within the range of 0.1 to 0.9.  
For RoBERTa and CN-CLIP, we use a two-layer MLP as the classification head. Specifically, for CN-CLIP, we concatenate its multi-modal representations as the input to the classifier.  
For YOLO, we set prompt token num as 30.  
For Intern-VL, we construct the SFT data using the same instruction format as the prompts for GPT-4o, directly treating the label as the output target.  
For all baselines, we follow their default settings or mainstream research when configuring training hyper-parameters, including batch size, learning rate, and other relevant settings. And all experiments are conducted on NVIDIA A800-SXM4-80GB machines.

\begin{table*}[ht]
    \centering
    \caption{Performance of Various Moderation Methods on the \M{} Benchmark. We categorize the moderation methods into \textbf{Binary Classification} and \textbf{Multi-class Classification}. The binary classification only determines whether a video is violative or not, while the multi-class classification requires the model to directly classify the video into its respective category. Optimal and sub-optimal performance is denoted in bold and underlined fonts, respectively.}
    \resizebox{1.0\textwidth}{!}{
    \begin{tabular}{l|cccc|cc|cc|c}
        \toprule
        \multirow{2}{*}{\textbf{Method}} & \multicolumn{4}{c|}{\textbf{Main Violative Category Recall}} & \multicolumn{2}{c|}{\textbf{Negative Content}} & \multicolumn{2}{c|}{\textbf{Positive Content}} & \textbf{Overall} \\
        
        & \small{\textbf{Law\&Safety}} & \small{\textbf{Content}} & \small{\textbf{Commercial}} & \small{\textbf{Intellectual}} & \textbf{Precision} & \textbf{Recall} & \textbf{Precision} & \textbf{Recall} & \textbf{Accuracy} \\
        \midrule
        \rowcolor{black!7}\multicolumn{10}{l}{\emph{\underline{Binary Classification}}} \\
        \textbf{$\text{Perspective}_{\text{toxicity}}$} & 0.6036 & 0.4598 & 0.3455 & 0.2941 & 0.6591 & 0.7059 & 0.5538 & 0.5000 & 0.6190 \\
        \textbf{$\text{Perspective}_{\text{severe}}$} & 0.1716 & 0.1092 & 0.0909 & 0.1765 & 0.5922 & \textbf{0.9170} & 0.5429 & 0.1351 & 0.5870 \\
        \textbf{GPT-4o mini} & 0.7235 & 0.6833 & 0.5091 & 0.5294 & 0.7679 & 0.7958 & 0.7057 & 0.6706 & 0.7430 \\
        \textbf{GPT-4o} & 0.8176 & 0.8101 & 0.8182 & 0.4706 & \underline{0.8346} & 0.7336 & 0.6870 & 0.8009 & 0.7620 \\
        \textbf{RoBERTa} & 0.8530 & 0.7667 & 0.7455 & {0.5882} & 0.6448 & 0.7915 & 0.6817 & 0.8174 & 0.7400 \\
        \textbf{CN-CLIP} & 0.7396 & 0.6034 & 0.7636 & {0.5882} & 0.7836 & 0.6777 & 0.7858 & 0.8633 & 0.7850 \\
        \textbf{YOLO} & 0.7929 & 0.5920 & 0.7818 & 0.4706 & 0.8189 & 0.6967 & 0.8003 & \underline{0.8875} & {0.8070} \\
        \textbf{Intern-VL-7B-SFT} & 0.8471 & 0.8056 & 0.8000 & \underline{0.6471} & 0.8152 & 0.7765 & \underline{0.8600} & 0.8287 & \underline{0.8230} \\
        \rowcolor{black!7}\multicolumn{10}{l}{\emph{\underline{Multi-Class Classification}}} \\
        \textbf{GPT-4o mini}  & \underline{0.8706} & 0.7889 & \textbf{0.8727} & 0.4118 & 0.6161 & 0.8175 & 0.8250 & 0.6280 & 0.7130 \\
        \textbf{GPT-4o} & {0.8647} & \underline{0.8278} & 0.8364 & 0.5294 & 0.6381 & 0.8318 & {0.8422} & 0.6557  & 0.7300 \\
        \textbf{Intern-VL-7B-SFT} & 0.6706 & 0.5944 &  0.7091 & {0.2353} &{ 0.7976} & 0.6256 & {0.7638} & 0.8841 & {0.7750} \\
        \textbf{\M{}-7B} & \textbf{0.8765} & \textbf{0.8333} & \underline{0.8545} & \textbf{0.7059} & \textbf{0.9701} & \underline{0.8460} & \textbf{0.8972} & \textbf{0.9810} & \textbf{0.9240} \\
        \bottomrule
    \end{tabular}
    }
    \label{tab:off_exp}
\end{table*}


\subsection{Evaluation Result}
As shown in \Cref{tab:off_exp}, Perspective API performs worse than other approaches, despite being a specialized moderation API. This is consistent with previous study \cite{moderation_detect_2}. We attribute \M{}'s performance to its relatively old version and inability to leverage visual information.  
In contrast, GPT-4o and GPT-4o mini are more aligned with the release dates of the videos and possess advanced multi-modal capabilities, leading to superior performance. This highlights the necessity of update frequency for industrial moderation APIs.


Methods trained on \M{} data, regardless of model size, consistently outperform API-based approaches. Even the RoBERTa model with only 500M parameters surpasses the large-scale GPT-4o, confirming our analysis about the importance of real-time updates for video moderation task. Among these methods, text-based models perform slightly worse than multi-modal models, indicating that while text provides useful information, certain critical information can only be effectively captured through complex visual content. 
Besides, Additionally, methods based on smaller models exhibit slightly lower performances compared to those based on larger models, as the limited capacity restricts their understanding ability.

When transitioning from binary classification to multi-class classification, both the untrained GPT-4o and the trained Intern-VL experience performance degradation, as multi-class classification is inherently more challenging than binary classification.  
In contrast, our \M{} approach achieves significantly better performance with multi-class training compared to the aforementioned baselines: It achieves the highest or second-highest \textit{precision}, \textit{recall}, and \textit{overall accuracy} for positive and negative content. Moreover, it demonstrates relatively balanced high recall across the four major violation categories, highlighting its strong discriminative capability and generalizability. We attribute this improvement to three key factors: the \Y{} foundation model, which is better suited for short video scenarios; the CoT state-transition data, which enhances \M{}'s content understanding; and the two-step offline training pipeline, which systematically refines moderation capabilities.  
\subsection{Analysis Experiments}

\subsubsection{Ablation study}
To validate the effectiveness of each component, we compare \M{} with its five variants:  
(1) Original YuanQi-7B model without post-training.
(2) Intern-VL-7B as the foundation model.
(3) Removing the Tag2CoT mechanism, directly generating moderation process with \Y{}.
(4) Removing the state transition workflow.
(5) Omitting the DPO training.
\begin{table}[h]
    \centering
    \caption{Ablation study on each component of \M{}.  
}
    \resizebox{0.8\columnwidth}{!}{
    \begin{tabular}{lccccc}
        \toprule
        \multirow{2}{*}{\textbf{Method}} & \multicolumn{2}{c}{{Negative Content}} & \multicolumn{2}{c}{{Positive Content}} &{Overall} \\
         \cmidrule(lr){2-3} \cmidrule(lr){4-5} \cmidrule(lr){6-6}
        & \textbf{Precision} & \textbf{Recall} & \textbf{Precision} & \textbf{Recall} & \textbf{Accuracy} \\
        \midrule
        {\M{}}  & \textbf{0.9701} & \textbf{0.8460}& \textbf{0.8972} & \textbf{0.9810} & \textbf{0.9240} \\
        \midrule
        YuanQi-7B & 0.4465 & 0.5047 & 0.6004 & 0.5433  & 0.5270 \\
        w Intern-VL & 0.9167 & \underline{0.8341} & \underline{0.8864}& 0.9446 & \underline{0.8980} \\
        w/o Tag2CoT & \underline{0.9530} & 0.7204 & 0.8267 & \underline{0.9740} & 0.8670 \\
        w/o state transition  & 0.9337 & 0.7346 & 0.8323 & 0.9619 & 0.8660 \\
        w/o DPO & 0.8834 & 0.8081 & 0.8681 & 0.9221 & 0.8740 \\
        \bottomrule
    \end{tabular}
    }
    \label{tab:ablation}
\end{table}

As shown in \Cref{tab:ablation}, the absence of any single component results in a performance drop: 
The native \Y{} model demonstrates moderate performance on the video moderation task, this validates the necessity of training adaption. While Intern-VL as foundation lags behind \Y{} in short video understanding.  
Removing Tag2CoT mechanism and the state transition workflow reduces \M{}'s ability to leverage video metadata.  
And DPO training further enhances the \M{}’s moderation capability beyond the improvements achieved through SFT.  



\section{ONLINE APPLICATIONS}
\label{sec: online_exp}
We explore the application of \M{} in both platform ecosystem governance and enhancing platform revenue on three scenarios of \C{}: NEBULA, Featured and Main Site. NEBULA is a lightweight platform fully focused on short video recommendations, the Featured platform pushes relatively higher quality short videos, while the Main Site encompasses the largest user base and \C{}'s business, including live streaming and e-commerce. Each of these scenarios has over 10 million daily active users (DAU), with more than 16 million short video impressions per day. 

\subsection{Deployment Details}

To improve response speed, we apply FP16 quantization to \M{}-7B model for deploying. On a cluster of 48 RTX4090 GPUs, KuaiMod achieves a QPS (Queries Per Second) of 6.81. This service moderates all candidate videos uploaded to the Kuaishou platform daily, with an average volume of 600,000 videos per day.

\subsection{Performance of A/B test}
\subsubsection{Comprehensive Ecosystem Governance}
In this aspect, \M{} provides its judgments for online videos, videos identified as violative are directly removed from promotion to refine platform ecosystem. To narrow the moderation scope, we select videos that have been reported more than five times as candidates to analyze their specific violations with \M{}. 

 The A/B test result in \Cref{tab:online_exp1} shows that \M{} significantly reduces the user reporting rate, indicating an improvement in user satisfaction. Furthermore, DAU and total App Usage Time (AUT) exhibited slight fluctuations or even an upward trend, highlighting the critical role of ecosystem governance in platform development.  
\begin{table}[h]
    \centering
    \begin{minipage}[b]{0.45\columnwidth}
        \centering
        \caption{\M{}'s online A/B test results for comprehensive ecosystem governance on \C{} NEBULA and Featured.}
        \resizebox{1.0\columnwidth}{!}{
            \begin{tabular}{lccc}
                \toprule
                \textbf{Column} & \textbf{Report Rate (\%) $\downarrow$} & \textbf{DAU (\%) $\uparrow$} & \textbf{Total AUT (\%) $\uparrow$} \\
                \midrule
                NEBULA & \cellcolor{green!30} -24.34 & \cellcolor{green!25} +0.016 & \cellcolor{red!5} -0.002 \\
                Featured & \cellcolor{green!20} -18.98 & \cellcolor{red!5} -0.002 & \cellcolor{green!25} +0.012 \\
                \bottomrule
            \end{tabular}
        }
        \label{tab:online_exp1}
    \end{minipage}
    \hfill
    \begin{minipage}[b]{0.52\columnwidth}
        \centering
        \caption{\M{}'s online A/B test result for personalized recommendation enhancement on \C{} Main Site.}
        \resizebox{1.0\columnwidth}{!}{
            \begin{tabular}{lccc}
                \toprule
                \textbf{Column} & \textbf{DAU (\%) $\uparrow$} & \textbf{Avg AUT (\%) $\uparrow$} & \textbf{Total AUT (\%) $\uparrow$} \\
                \midrule
                Main Site & \cellcolor{green!20} +0.028 & \cellcolor{green!20} +0.033 & \cellcolor{green!30} +0.063 \\
                \bottomrule
            \end{tabular}
        }
        \label{tab:online_exp2}
    \end{minipage}
\end{table}

\subsubsection{Personalized Recommendation Enhancement}
\label{recommend}
In this aspect, \M{}’s judgments are embedded into features to enhance personalized video recommendation, which leads to personalized moderation strategy. The personalized moderation strategy is crucial because different users may have varying perceptions of harmful content. For example, content suitable for adults might be inappropriate for minors. However, only certain types of violating videos are preserved for personalized recommendation;
videos that explicitly violate laws or pose severe harm are strictly filtered. 

As shown in \Cref{tab:online_exp2}, both DAU and AUT increase due to the personalized moderation strategies, the total consumption time increases by 0.063\%.  
This indicates the potential of the fine-grained harmful content governance for improving recommendation revenue. In the future, we will further develop new paradigms for using VLMs in short video recommendations based on \M{}.

\section{CONCLUSION}
\label{sec: conclusion}

In this paper, we first analyze the challenges faced by traditional video moderation in industrial SVP, including the mental toll and high cost of manual review, human biases. Then evaluate existing automated moderation methods with self-constructed \M{} benchmark to declare  their limitations in complex content understanding, user feedback perception and lengthy update cycle. 
To address these issues, we propose \M{}, a content ecosystem governance framework for SVP inspired by common-law system. Unlike traditional approaches, \M{} defines violations based on concrete cases rather than intricate regulations. 
\M{} first unlock the potential of VLMs for video moderation with CoT and state-transition data. Then it adopts RL paradigm to continuously refine moderation policy through interactions with the online environment. \M{} outperforms all baselines on offline benchmarks and significantly improves user experience  after online deployment. Our experience demonstrates the crucial role of ecosystem governance in the development of SVP.

\bibliographystyle{plain} 
\bibliography{main}
\clearpage

\appendix

\setcounter{section}{0}
\setcounter{figure}{0}

\renewcommand\thesection{\Alph{section}}
\renewcommand {\thefigure} {A-\arabic{figure}}

\renewcommand\thesubsection{\Alph{section}.\arabic{subsection}}
\section{APPENDIX}
\subsection{Instructions for Different Tasks}
In this section, we present all the instructions to process data during the development of \M{}, as well as some prompts used to implement the baseline model.
\subsubsection{Instruction for Tag2CoT data generation process.}
In \Cref{fig:tag2cot}, we present the instruction to guide the YuanQi model in conducting causal analysis based on video metadata and annotation results, and generating rationales for violation labels. This instruction is used the \textbf{Tag2CoT} process in \Cref{subsec: workflow}.

\subsubsection{Instruction for CoT2Tag data generation process.}
In \Cref{fig:cot2tag}, we present the instruction used the \textbf{CoT2Tag} process to prompt \Y{}-20B model to organize the analysis process into our moderation workflow in the state-transition format introduced in \Cref{subsec: workflow}.
\subsubsection{Instructions to Prompt GPT-4o and Construct Intern-VL SFT Data.}
\begin{figure}[ht]
    \centering
    \begin{subfigure}[b]{0.48\columnwidth}
        \centering
        \includegraphics[width=\textwidth, height=7.5cm]{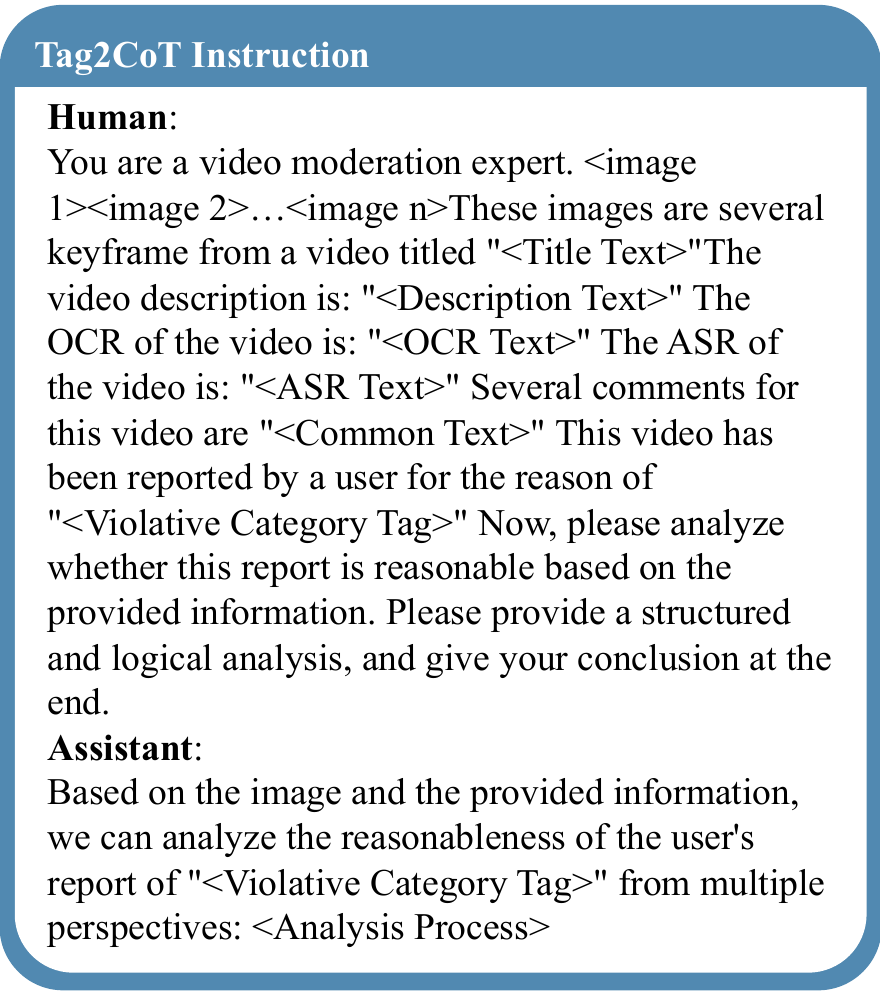}
        \caption{Instruction for \Y{}-20B to generate CoT rationales for video's violation category tag based on video metadata and use comment.}
        \label{fig:tag2cot}
    \end{subfigure}
    \hfill
    \begin{subfigure}[b]{0.48\columnwidth}
        \centering
        \includegraphics[width=\textwidth, height=8cm]{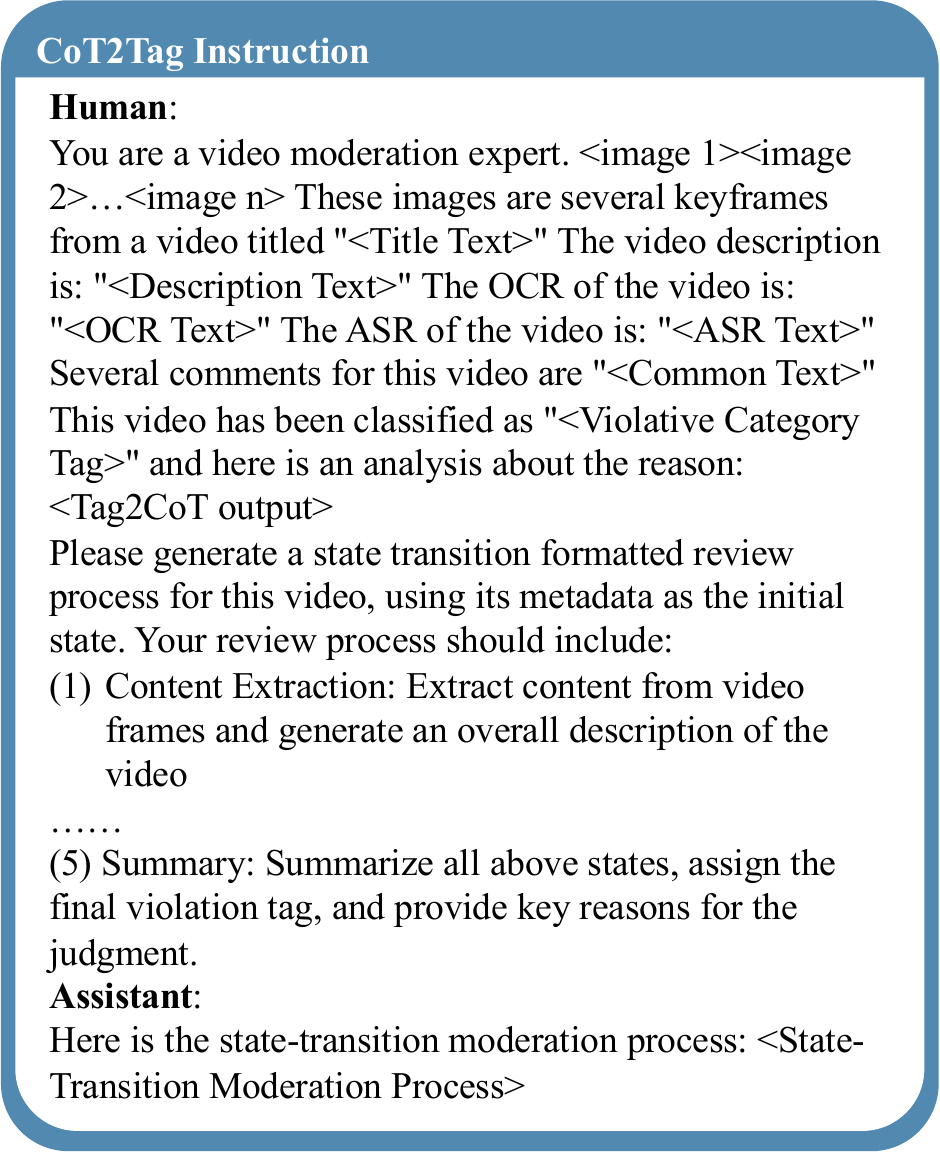}
        \caption{Instruction for \Y{}-20B to organize metadata, analysis and violative tag into a workflow.}
        \label{fig:cot2tag}
    \end{subfigure}
    \vskip\baselineskip
    \begin{subfigure}[b]{0.48\columnwidth}
        \centering
        \includegraphics[width=\textwidth, height=4cm]{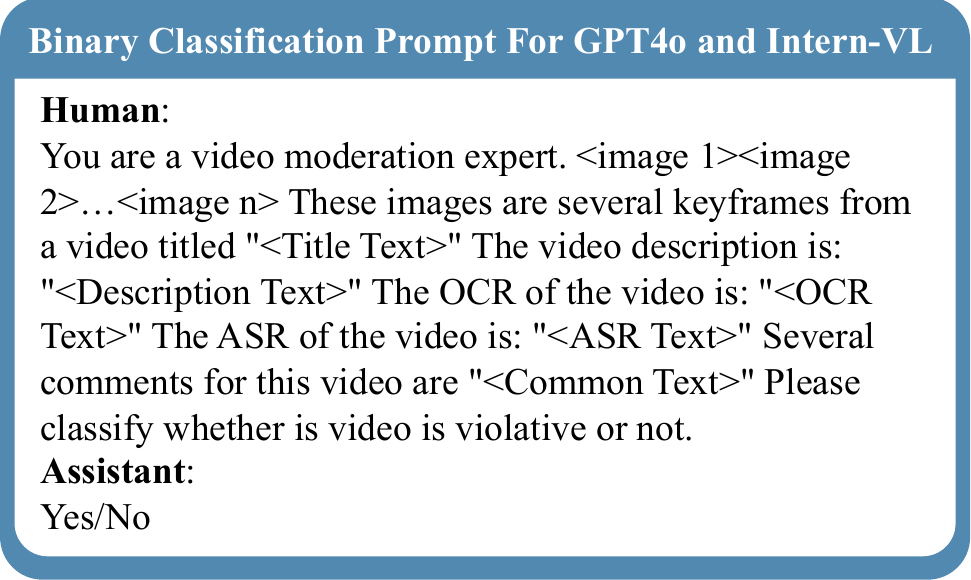}
        \caption{Instructions for binary-class classification.}
        \label{fig:prompt1}
    \end{subfigure}
    \hfill
    \begin{subfigure}[b]{0.48\columnwidth}
        \centering
        \includegraphics[width=\textwidth, height=4cm]{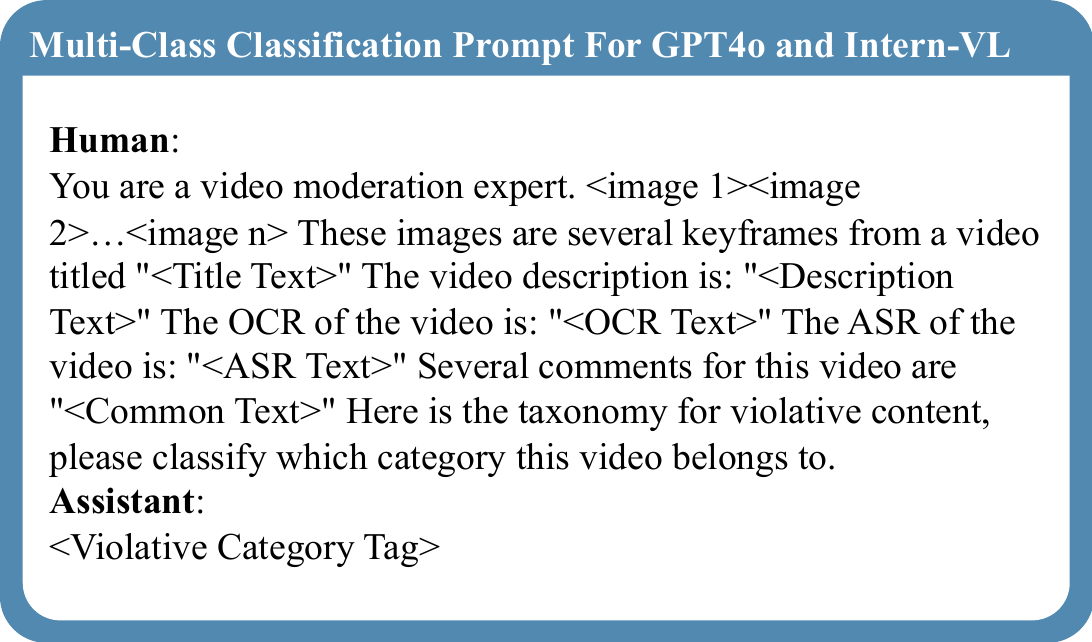}
        \caption{Instructions for multi-class classification.}
        \label{fig:prompt2}
    \end{subfigure}
    \caption{Instructions for \Y{} on different tasks.}
    \label{fig:prompts}
\end{figure}

In \Cref{fig:prompt1} ans \Cref{fig:prompt2}, we present the instruction used in offline evaluation experiments (\Cref{sec: experiments}) to prompt GPT-4o mini, GPT-4o and to construct SFT data for Intern-VL model.
\label{sec:instructions}



\subsection{Case Study}
\label{sec:case_study}
\begin{figure}[ht]
    \centering
    \begin{subfigure}[b]{0.48\textwidth}
        \centering
        \includegraphics[width=\textwidth, height=8cm]{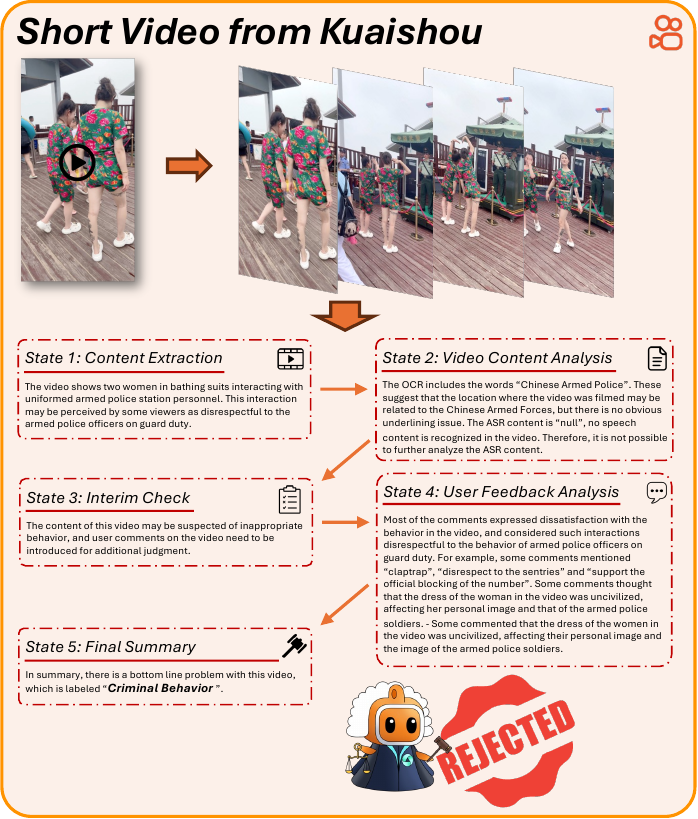}
        \caption{Violative case 1 of \M{}.}
        \label{fig:case1}
    \end{subfigure}
    \hfill
    \begin{subfigure}[b]{0.48\textwidth}
        \centering
        \includegraphics[width=\textwidth, height=8cm]{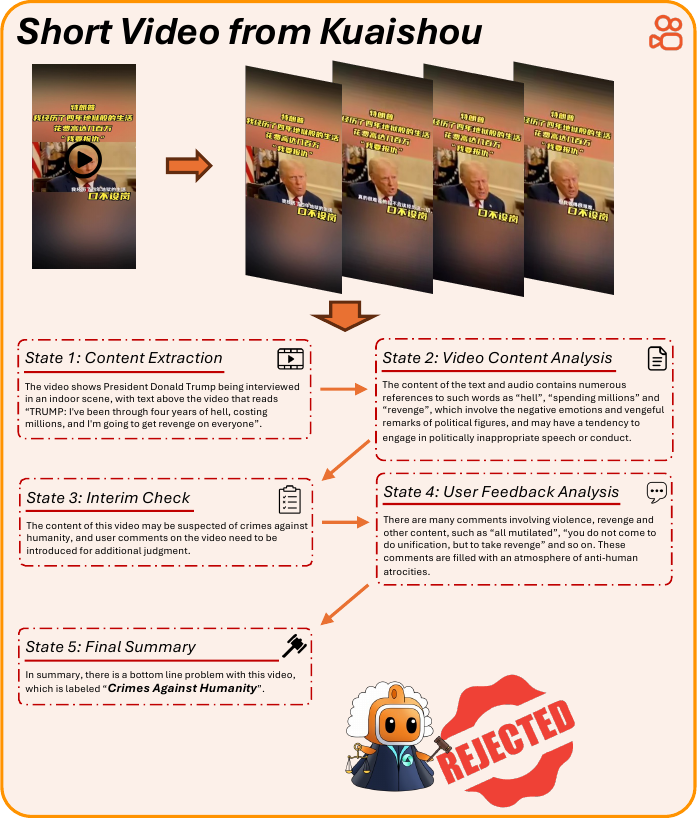}
        \caption{Violative case 2 of \M{}.}
        \label{fig:case2}
    \end{subfigure}
    \vskip\baselineskip
    \begin{subfigure}[b]{0.48\textwidth}
        \centering
        \includegraphics[width=\textwidth, height=8cm]{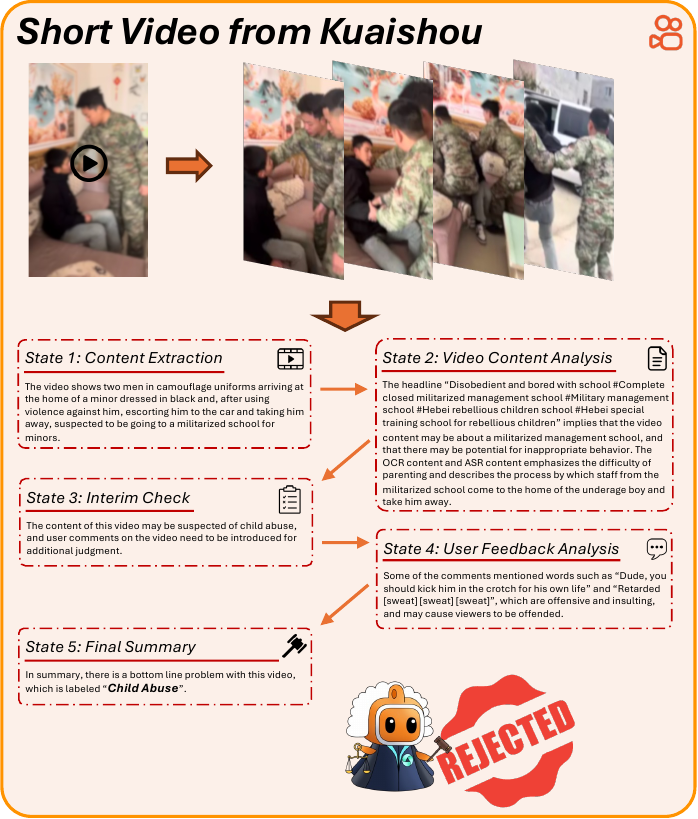}
        \caption{Violative case 3 of \M{}.}
        \label{fig:case3}
    \end{subfigure}
    \hfill
    \begin{subfigure}[b]{0.48\textwidth}
        \centering
        \includegraphics[width=\textwidth, height=8cm]{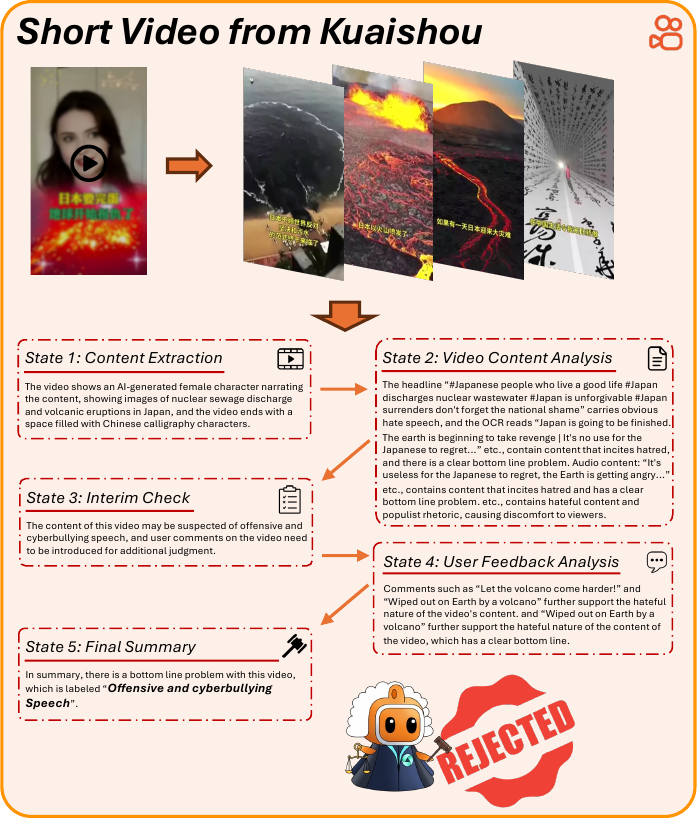}
        \caption{Violative case 4 of \M{}.}
        \label{fig:case4}
    \end{subfigure}
    \caption{Violative cases 1 to 4 of \M{}.}
    \label{fig:cases}
\end{figure}

\begin{figure}[ht]
    \centering
    \begin{subfigure}[b]{0.48\textwidth}
        \centering
        \includegraphics[width=\textwidth, height=8cm]{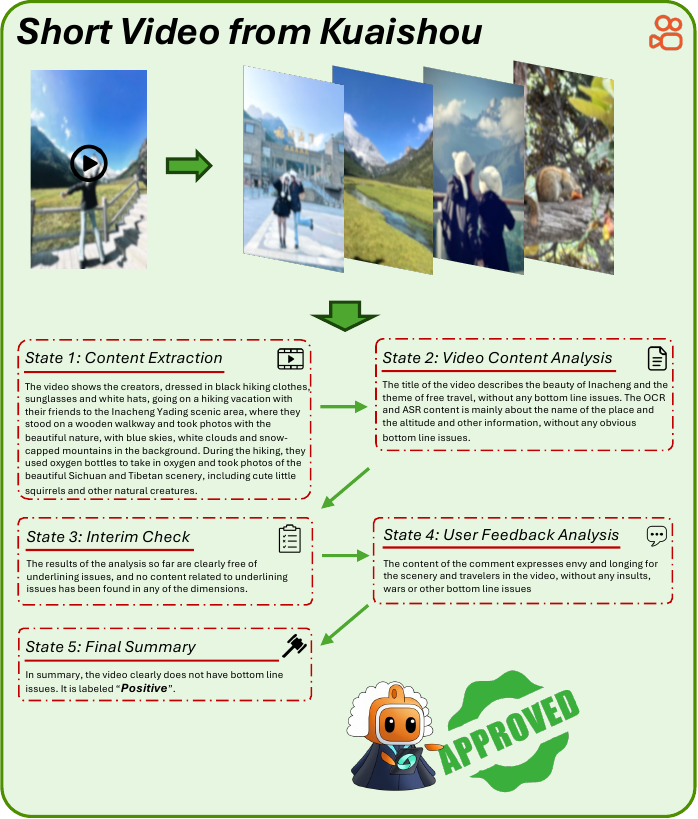}
        \caption{Positive case 1 of \M{}.}
        \label{fig:case5}
    \end{subfigure}
    \hfill
    \begin{subfigure}[b]{0.48\textwidth}
        \centering
        \includegraphics[width=\textwidth, height=8cm]{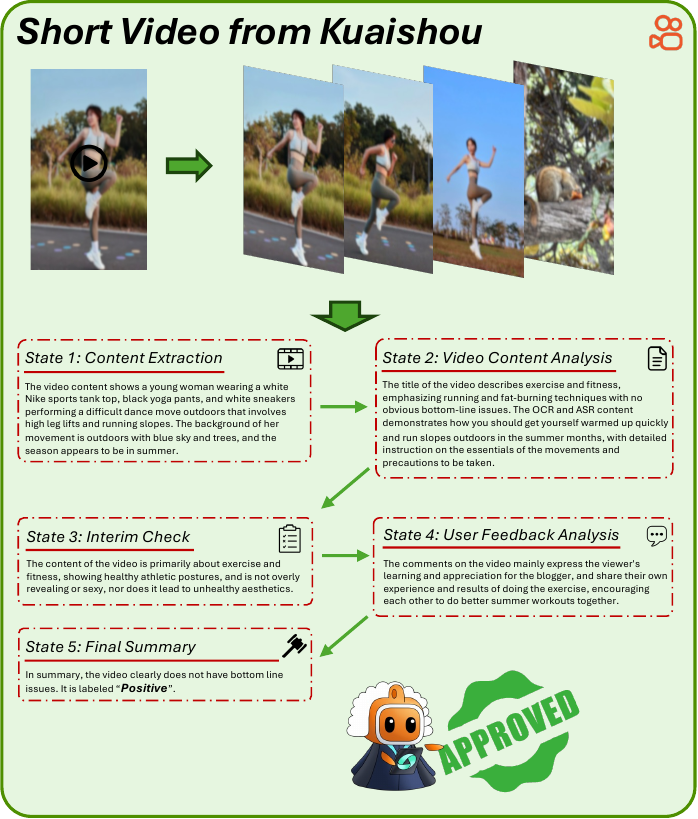}
        \caption{Positive case 2 of \M{}.}
        \label{fig:case6}
    \end{subfigure}
    \caption{Positive cases 1 to 2 of \M{}.}
    \label{fig:cases5and6}
\end{figure}

In this section, as shown in \Cref{fig:case1,fig:case2,fig:case3,fig:case4,fig:case5,fig:case6} , we present several real-world case analyses and decision results of \M{}, drawn from live online examples. These cases are deliberately challenging and potentially misleading, demonstrating \M{}'s superior moderation capabilities and providing an intuitive understanding of the data format and task definitions. In the input context, we provided multi-frame images of videos, titles, OCR text, ASR text, and the top 20 hottest user comments in the comment section of that video. \M{} performs step-by-step state analysis and transfer based on the multimodal information of these videos to obtain the final conclusion. From these examples, we can see that \M{} can effectively help the \C{} platform identify undesirable content, such as criminal behavior, offensive and cyberbullying speech, and other illegal platform bottom lines, while protecting more quality creators and videos and creating a safer and healthier community atmosphere.
\subsection{Ethics Statement}
This study adheres to strict ethical standards to ensure the credibility and confidentiality of the findings. All data underwent a thorough de-identification process to protect the privacy and maintain anonymity of the users of the short video platform. The study follows ethical guidelines and informed consent was obtained from the participants while prioritizing their rights and autonomy. Transparency and accountability are maintained throughout the study to minimize bias and conflict of interest. No academic ethical issues or misconduct are encountered and the authors affirm their unwavering commitment to upholding ethical research practices and addressing any unintentional errors or omissions in a timely manner.

\clearpage

\clearpage

\end{document}